\documentclass[structabstract]{aa}  
\usepackage{graphicx}
\usepackage{txfonts}
\usepackage{natbib}
\bibpunct{(}{)}{;}{a}{}{,} 
\usepackage{color}

\begin{document}
   \title{A deep wide-field sub-mm survey of the Carina Nebula complex\thanks{Based on data acquired with the Atacama Pathfinder Experiment (APEX). APEX is a collaboration between the Max-Planck-Institut f\"ur Radioastronomie, the European Southern Observatory, and the Onsala Space Observatory.}}

   \author{Thomas Preibisch \inst{1} \and Frederic Schuller\inst{2}
          \and Henrike Ohlendorf\inst{1} \and Stephanie Pekruhl\inst{1}
          \and Karl M.~Menten\inst{2} \and Hans Zinnecker\inst{3} }

   \institute{
              Universit\"ats-Sternwarte M\"unchen, 
              Ludwig-Maximilians-Universit\"at,
              Scheinerstr.~1, 81679 M\"unchen, Germany\,\,
              \hfill \email{preibisch@usm.uni-muenchen.de}
         \and
             Max-Planck-Institut f\"ur Radioastronomie,
             Auf dem H\"ugel 69, 53121 Bonn, Germany   
        \and
             Astrophysikalisches Institut Potsdam, An der 
             Sternwarte 16, 14482 Potsdam, Germany
             }

\titlerunning{Deep wide-field sub-mm survey of the Carina Nebula}
\authorrunning{Preibisch et al.}

   \date{} 

 
  \abstract
   {
The Great Nebula in Carina is one of the most massive 
($M_{\ast,{\rm total}} \ga 25\,000\,M_\odot$) star-forming
complexes in our Galaxy and contains several stars with
(initial) masses exceeding $\approx 100\,M_\odot$; it is therefore
a superb location 
in which to study the physics of violent massive star-formation 
and the resulting feedback effects, 
including cloud dispersal and triggered star-formation.
   }
   {
{We aim to reveal 
the cold dusty clouds in the Carina Nebula complex, to
determine their morphology and masses, and to study the interaction
of the luminous massive stars with these clouds}.
   }
   {
We used the Large APEX Bolometer Camera LABOCA at the APEX telescope to map a 
$1.25\degr \times 1.25\degr$ ($\cor 50 \times 50$~pc$^2$) region
at $870\,\mu$m with
$18''$ angular resolution ($=$ 0.2~pc at the distance of the Carina Nebula) 
and an r.m.s.~noise 
level of $\approx 20$~mJy/beam.
   }
   {
From a comparison to H$\alpha$ images we infer that
about $6\%$ of the $870\,\mu$m flux in the observed area is likely
free-free emission from the H{II} region, while 
about $94\%$ of the flux is very likely
thermal dust emission.
The total (dust + gas) mass of all clouds for which
our map is sensitive is  $\sim 60\,000\,M_\odot$,
in good agreement with the mass of the compact
clouds in this region derived from $^{13}$CO line observations.
There is a wide range of different cloud morphologies and sizes,
from large, massive clouds with several $1000\,M_\odot$,
to small diffuse clouds containing just a few $M_\odot$.
We generally find good agreement in the 
cloud morphology seen at $870\,\mu$m and 
the Spitzer $8\,\mu$m emission maps, but also identify
a prominent infrared dark cloud.
Finally, we construct a
radiative transfer model for the Carina Nebula complex 
that reproduces the
observed integrated spectral energy distribution reasonably well.
   }
   {Our analysis suggests a total gas $+$ dust mass of about $200\,000\,M_\odot$
in the investigated area;
most of this material is in the form of molecular clouds, but
a widely distributed component of (partly)
atomic gas, containing up to $\sim 50\%$ of the total mass, may also
be present.
Currently, only some 10\% of the gas is in sufficiently dense clouds to be 
immediately available for future star formation, but this fraction may increase
with time owing to the ongoing compression of the strongly irradiated
clouds and the expected 
shockwaves of the imminent supernova explosions.
}

   \keywords{Stars: formation -- ISM: clouds -- ISM: structure -- 
  ISM: individual objects: \object{NGC 3372}  -- Submillimeter: ISM
               }

   \maketitle
%

\section{Introduction}

Most stars in the Galaxy are born in massive star-forming regions 
\citep[e.g.,][]{Briceno07}.
The massive stars 
profoundly influence their environments by creating H\,{\sc II}
regions, generating 
wind-blown bubbles, and exploding 
as supernovae. This feedback disperses the natal molecular clouds
\citep[e.g.,][]{Freyer03}, thus in
principle halting further star formation, but ionization fronts and expanding 
superbubbles can also compress nearby clouds and may thereby trigger the 
formation of new generations of stars \citep[e.g.,][]{Dale05,Dale08,Gritschneder09}. 
These processes determine the key outputs from star-formation,
such as the stellar mass function, 
the total star formation
efficiency, and the frequency of planetary formation. 
Detailed observational diagnostics of these processes
have been hard to come by, mainly
because regions with large populations of massive stars and strong feedback 
are usually too far away for detailed studies. 

At a distance of 2.3~kpc, the Carina Nebula 
\citep[NGC 3372; see, e.g.,][for an overview]{SB08}
is the nearest 
southern region with a large massive stellar population 
\citep[65 O-type stars; see][]{Smith06}, including several of the
most massive and luminous stars known in our Galaxy.
The Carina Nebula complex (CNC hereafter) has a total infrared luminosity
of $\sim 10^7\,L_\odot$ and a size of about 50 pc,
corresponding to an extent of $1.25\degr$ on the sky.
Most of the massive stars reside in one of several
clusters \citep[Tr~16, Tr~14, Tr~15; see][]{Trumpler} with ages ranging from 
$\sim 1$ to several~Myr.
 In the central region around $\eta$~Car and the 
Tr~16 cluster, the molecular clouds have already been largely 
dispersed by stellar feedback. 
In the South Pillars 
(southeast of $\eta$~Car) the clouds are
eroded and shaped by the radiation and winds from
$\eta$~Car and Tr~16, giving
rise to numerous giant dust pillars, which feature very prominently 
in the mid-infrared images made with the {\it Spitzer} Space Observatory
\citep{Smith10b}. The detection of
young stellar objects \citep{Megeath96} and a very young cluster
\citep[the ``Treasure Chest Cluster''; see][]{Smith05}
embedded within the heads of some of the dust pillars
suggests
that the formation of a new generation of stars is currently triggered in this area.

Because of its prominence on the sky, the CNC 
 has been the target of numerous observations in almost
all wavelength bands from $\gamma$-rays to the radio regime
\citep[see][for a summary]{SB08}.
Recent deep large-scale 
surveys of the CNC have been obtained with the
the {\it Hubble} Space Telescope (HST) \citep{Smith10a}, 
the {\it Spitzer} infrared observatory \citep{Smith10b}, 
the near-infrared camera HAWK-I at the ESO Very Large Telescope
\citep{Preibisch11a}, and
the {\it Chandra}  X-ray observatory \citep{Townsley11,Preibisch11b}.
All these together will allow us for the first time to identify the young 
low-mass star population in the complex without the strong 
confusion problems that plague studies based only on optical and 
infrared data sets.

However, any
really comprehensive investigation of this region  clearly
also requires information on the 
cool dust and gas in the (molecular) clouds, and the deeply 
embedded protostars within these clouds.
During the very earliest stages of star formation,
these dense gas clumps and cores remain very cold
($10-30$~K), and therefore escape detection at near- and
mid-infrared wavelengths, even with instruments as sensitive
as {\it Spitzer}.
Only the (sub-)\,millimeter and radio
emission from molecular spectral lines and
from cool dust  allows for an almost un-hindered, unique view onto
the processes in the dense clouds.
In order to meaningfully
complement the extraordinary quality of the recent HST, {\it Spitzer},
{\it Chandra}, and HAWK-I data, (sub-)mm 
observations with high spatial resolution,
high sensitivity, and
large spatial coverage (at least 1 square-degree) 
are therefore clearly required.

Until recently, the
best existing mm-band data set was the SEST/SIMBA survey
of the central region of the Carina Nebula by
\citet{Brooks05}.
Their map of the 1.2~mm continuum emission covers an area of
$10' \times 10'$ with a half-power beam width of $24''$;
with a sensitivity limit of 75~mJy per beam,
clumps with masses down to
$\sim 5\,M_\odot$ were detected in this survey.
\cite{Gomez10} recently studied the central $30' \times 30'$ area
of the CNC, covering $\eta$~Car and the Keyhole Nebula,
 at $870\,\mu$m with LABOCA at the APEX telescope.
At radio wavelengths, the most comprehensive existing data set
is a NANTEN survey in several CO lines, covering
a $4\degr \times 2\degr$ area  
with a half-power beam width of $2.7'$ \citep{Yonekura05}.
The Mopra $^{12}$CO (1--0) data with a $43''$ beam presented by
\citet{Brooks98} \citep[see also][]{Schneider04}
provide better spatial resolution,
but cover only a small area in the central part of the CNC.
There was therefore an obvious need for (sub-)mm-data with high spatial 
resolution as well as high sensitivity, that cover a
considerably larger part of the CNC.

In this paper we present the results of our large-scale sub-mm
mapping of the CNC with LABOCA at the APEX telescope.
 Our map covers a $\sim 6$ times
larger area and is about twice as sensitive as the
sub-mm~data from \cite{Gomez10}.
 After the description of the observations and
data analysis (Sect.~2), we discuss general aspects (Sect.~3),
describe the structure and properties of the clouds in different regions
within the complex (Sect.~4) and compare them to recent {\it Spitzer} maps
(Sect.~5). In Sect.~6 we take a look at the sub-mm properties of
dense molecular clumps found in CO line observations, and in Sect.~7
we consider global properties of the complex, such as the total cloud
mass, and perform radiative transfer modeling of the integrated 
spectral energy distribution to derive a global model for the CNC.


\section{Observations and data analysis}

   \begin{figure*} \sidecaption
   \includegraphics[width=12cm]{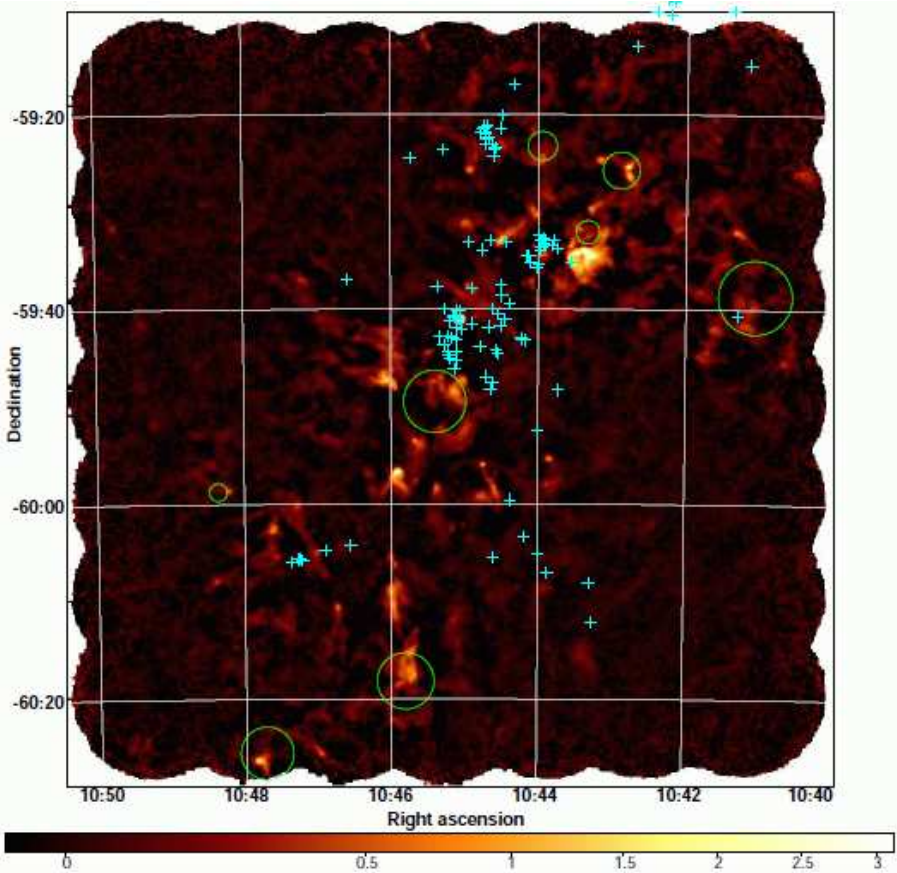}
   \caption{
False-color representation of the wide-field LABOCA map of the
Carina Nebula complex. 
The cyan crosses mark the positions of the massive (O- and early B-type) stellar
members of the Carina Nebula from \cite{Smith06},
the green circles mark the locations and radii of the dense
C$^{18}$O clumps detected by \cite{Yonekura05} in this area.
The field-of-view is $\approx 1.3\degr \times 1.3\degr$, corresponding to
a size of $\approx 52 \times 52$~pc at the distance of the complex; 
the grid shows the J2000 coordinates.
 The units of the scale bar at the bottom are
Jy/beam.\label{laboca-map}}
    \end{figure*}

The sub-mm observations of the CNC presented here
were performed  with
the ``Atacama Pathfinder Experiment'' (APEX),
a 12-meter radio telescope  in Chile's Atacama
desert, the best accessible site for submillimeter observations.
The  APEX telescope is described in detail in \citet{Guesten06}.
We used the Large Apex BOlometer CAmera LABOCA \citep[see][]{Siringo09}, which
operates
in the atmospheric window at $870\,\mu$m (345 GHz). It has 295 pixels
arranged in a hexagonal layout consisting of a center channel
 and nine concentric hexagons. 
The angular resolution is $18.6''$, and the
 total field of view for LABOCA 
is $11.4'$, making it an ideal instrument for mapping large
regions with high sensitivity.
At the 2.3~kpc distance of the Carina Nebula, its angular 
resolution corresponds to a linear dimension
of 0.2~pc. This is sufficient to resolve the structure
of molecular {\em clumps} (i.e.~relatively large dense clouds linked to the
formation of small stellar clusters), but not the individual cloud 
{\em cores} (i.e.~very compact clouds out of which individual stellar
systems form), which have typical sizes of $\sim 0.1$~pc or less.

The LABOCA observations of the CNC discussed here
were obtained on 22, 24 and 26 December 2007.
For the mapping we employed the ``raster map in spiral mode''.
The total area to be mapped was covered with a raster of pointings similar 
to, but larger than the one shown in Fig.~9 of \citet{Siringo09}. 
At each pointing, fully sampled maps
 of the total field-of-view of LABOCA were obtained by moving the
telescope along a spiral pattern.
The total on-source integration time used for our mapping of the CNC
was $\approx 10$~hours.
The observing conditions were good, with a precipitable water vapor column
of $< 2$~mm.

The data were reduced with the BOlometer array Analysis software
(BOA) package, following
the procedures described in detail in
\citet{Schuller09}. As the final product of the data analysis,
a map  with a pixel size of $6.07''$ (i.e.~$\sim 3$~pixels per beam)
was compiled. With a LABOCA beamsize of 392 square-arcseconds,
the pixel-to-beam-size ratio for the transformation from
surface brightness to integrated fluxes is  0.0941~beams/pixel.

The data were calibrated by applying an
opacity correction, as determined from skydips observed typically
every two hours \citep[see][]{Siringo09}.
In addition, the flux calibration was regularly 
checked against primary calibrators (planets) 
or secondary calibrators (bright Galactic sources).
The total calibration error should be lower than 15\%.

A fundamental limitation comes from the removal of correlated noise
in the data reduction; because of this, our map is not sensitive to
any structures with angular sizes larger than the array ($\ge 10'$)
and can only partly recover emission on scales larger than 
$\approx 2.5'$.
As a consequence, possible {\em uniform}
emission on angular scales $\ga 2.5'$ is filtered out and thus
absent in our map. Therefore, the fluxes we measure in our map
are always lower limits to the true sub-mm sky fluxes.
The amount of this unseen flux depends on the (unknown)
spatial distribution of the total emission in the map.
If most of the emission comes from well localized, dense clouds,
the missing flux will be very low or even negligible;
the fluxes from individual compact ($< 2.5'$) clouds can be reliably 
determined.
If, on the other hand, there is bright large-scale emission,
the measured flux can be considerably lower than 
the true total flux.
These issues will be discussed in more detail in Sec.~\ref{total_mass_sec}.

\section{General results}

The result of our LABOCA observations 
is a map of the $870\,\mu$m emission,
covering a total area of $1.25\degr \times 1.25\degr$; it is
shown in Fig.~\ref{laboca-map}.
The r.m.s.~noise level in the map is $\approx 20$~mJy/beam.
For isolated compact clumps with assumed uniform temperatures of
$T \approx 20-30$~K, this corresponds to a nominal
sensitivity limit for the clump masses of $\sim 2\,M_\odot$.
This map provides the
{\em first spatially complete survey of the sub-mm emission in the CNC}.

The maximum intensity in our map, 31.5~Jy/beam, is found
at the position of $\eta$~Car.
Emission from clouds is seen with intensities up to
$\approx 4$~Jy/beam. 
From these maximum cloud fluxes 
we can immediately infer \citep[see, e.g., Eq.~3 in][]{Schuller09}
that
the optical depth of the emitting clouds in our map is  small,
$\tau_{870\,\mu\rm m} < 0.01$, i.e.~all observed
cloud emission is clearly in the 
optically thin limit.

In order to determine the total sub-mm flux in  our LABOCA map,
we excluded 25 pixel-wide edges of the mosaic, where the pixel
values are often dominated by noise, and  integrated all
pixel values exceeding the $3\sigma$ noise level of 60~mJy/beam.
We found a total $870\,\mu$m flux of 1147~Jy for our map.

\subsection{Nature of the observed sub-mm emission \label{emission_nature.sec}}

   \begin{figure*}\begin{center}
   \includegraphics[width=16cm]{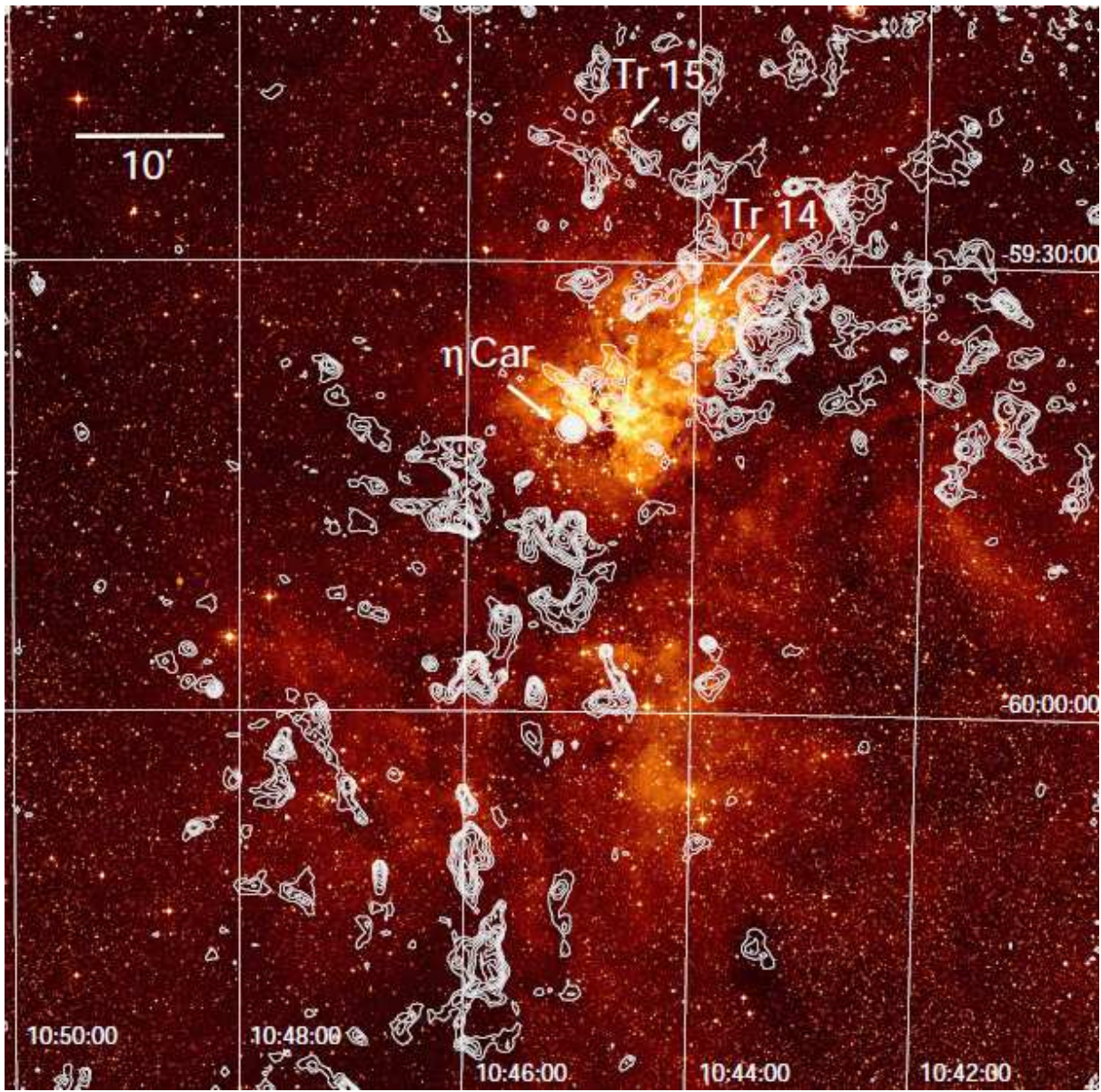} \end{center}
   \caption{False-color representation of the red optical
Digitized Sky Survey 
image with contours of the LABOCA map overplotted.
Here and (unless noted otherwise) in the following images, 
the first three contour levels are 0.06, 0.12, and 0.18~Jy/beam,
while the further levels increase by a factor of $\sqrt{2}$. 
The grid shows the J2000 coordinates.
              \label{laboca-dss}%
    }
    \end{figure*}

Two fundamentally different emission mechanisms  can
contribute to the observed $870\,\mu$m continuum emission:
thermal emission from dust, and free-free emission from ionized gas.
The second mechanism is relevant in some parts of the CNC,
where the numerous massive stars
produce an extremely strong ionizing radiation field
and create extended stellar wind bubbles.
The H$\alpha$ line is a good and easily detectable tracer of
ionized gas.
To localize H$\alpha$ emission in the region we retrieved
from the ESO archive images obtained with the Wide Field Imager
at the MPI-2.2~m Telescope through an H$\alpha$ filter
($\lambda_{\rm center} = 658.827$~nm, $FWHM = 7.431$~nm)
for inspection.
We also used the publicly available
High Level Science Products of the
HST ACS H$\alpha$ Survey of the Carina Nebula\footnote{The
data set and the processing are described at
{\tt http://archive.stsci.edu/pub/hlsp/carina/\newline
hlsp\_carina\_hst\_v2\_readme.txt}},
Version 21 December 2009, created by \cite{Mutchler09}.
The diffuse emission visible in the optical red DSS image
(see Fig.~\ref{laboca-dss}) closely follows the H$\alpha$ emission.

While these optical images trace the ionized gas,
the distribution of the cold molecular gas can be inferred from
molecular line maps. 
A comparison of the optical images with the CO maps presented in
\citet{Yonekura05} shows a fairly clear spatial anti-correlation between 
the ionized and the molecular gas in the CNC
\citep[see also Fig.~3 in][{[SB07 hereafter]}]{SB07}. 
Strong H$\alpha$ emission is predominantly concentrated
in the very center of the CNC, close to $\eta$~Car and the hot 
stars in Tr~16, as well as near the massive cluster Tr~14. 
The CO emission, on the other hand, is concentrated in two large
cloud structures, approximately to the north and south of $\eta$~Car.

Our comparison of the LABOCA map to the red optical image
in Fig.~\ref{laboca-dss} confirms this anti-correlation.
Most of the sub-mm emitting clouds are located in regions
more than a few arcminutes away from $\eta$~Car,
where no (or at most very weak) H$\alpha$ emission can be seen,
and often
correlate well with the molecular emission. In these regions,
the sub-mm emission
is dominated by thermal emission from cool dust.
However, we also find some sub-mm emission in the central parts,
very close to $\eta$~Car or Tr~14. In these regions,
the sub-mm emission is 
likely to be strongly contaminated and probably dominated
by  free-free emission, instead of thermal dust emission.

A quantitative estimate of the contributions of these
different emission processes to the observed total  sub-mm flux
can be made in the following way: 
The total flux in the central $6' \times 7'$ area, which
contains $\eta$~Car and the Keyhole Nebula and where
most of the flux is likely free-free emission, is 70~Jy.
The total flux in the whole map excluding this central
area is 1077~Jy. Thus we conclude that about
94\% of the observed sub-mm emission in our map comes from dusty clouds,
whereas some 6\% of the total flux is probably free-free
emission.

\subsection{Fluxes and cloud masses\label{flux-to-mass}}

The standard way to determine cloud masses from observed 
(sub-)mm fluxes for optically thin thermal dust
emission is via a temperature-dependent scaling factor 
\citep[see][]{Hildebrand83}, by using the formula 
\begin{equation} 
M = \frac{D^2\,F_{\nu}\,R}{B_{\nu}(T_d)\;\kappa_{\nu}}\;\; ,
\end{equation}
where
$D$ is the distance to the source (2300~pc in our case), 
$F_{\nu}$ is the observed spectral flux density,
$R$ is the gas-to-dust mass ratio, 
$B_{\nu}(T_d)$ is the blackbody spectral flux density
for a dust temperature $T_d$, 
and $\kappa_{\nu}$κis the dust emissivity.
Following \citet{Schuller09}, we assume a
gas-to-dust mass ratio of $R = 100$ and a dust emissivity
of $\kappa_{870\,\mu{\rm m}} = 1.85\,{\rm cm}^2\,{\rm g}^{-1}$.
The dust emissivity depends on the detailed properties
of the dust grains; the value we use here is representative
for relatively dense molecular clouds \citep{Ossenkopf94,Henning95},
but deviations by about a factor of 2 cannot be excluded.

The most important factor of uncertainty is the choice of the
temperature.
Firstly, the individual clouds in the CNC will have 
different temperatures, depending on (a) their location with 
respect to the luminous massive stars and on the cloud density,
and (b) whether or not a cloud contains embedded protostars 
that heat it from inside. 
For some of the clouds in the CNC, temperature measurements are available:
\citet{Brooks03} found CO excitation
temperatures of $20 - 30$~K for the dense clouds near Tr~14,
and \citet{Yonekura05} derived mean $^{12}$CO excitation temperatures of
$20-23$~K for the large scale clouds and C$^{18}$O excitation temperatures
between 9~K and 28~K for dense clumps in the CNC.
Owing to the quite strong and non-linear dependence of the mass estimates 
on the assumed cloud temperature\footnote{For example, a
flux of 0.25~Jy corresponds to a cloud mass of
$23\,M_\odot$ for $T=10$~K, $7.3\,M_\odot$ for $T=20$~K, and
$4.2\,M_\odot$ for $T=30$~K.},
we have to consider the individual conditions in each cloud
in order to derive meaningful  mass estimates.
Secondly, even for an individual compact cloud, the
often used assumption of a spatially uniform ``characteristic'' temperature
is probably not correct.
A constant temperature  may appear to be a reasonable approximation for 
dense clumps in nearby low-mass star-forming regions (where the external 
heating of the clouds is weak), but
detailed modeling has shown that even in these relatively simple
cases,
slight deviations from isothermality by just a few degrees
can easily lead to errors in the estimated clump masses
by factors of $\sim 2$ \citep{Stamatellos07}.
In the harsh environment of the CNC, where the
level of cloud irradiation and heating by nearby hot and luminous 
stars is orders of magnitude higher than in quiescent regions,
the assumption of a constant cloud temperature in the fairly large
observed clumps can thus hardly be correct.

To summarize, the uncertainties about the dust opacities and  the
cloud temperatures will cause uncertainties of (at least)
about a factor of $\sim 2-3$ for the mass estimates.


\subsection{Cloud column densities}

As discussed in detail in Sec.~\ref{emission_nature.sec},  the
sub-mm flux we observe in most (but not all) parts of our map is most
likely thermal dust emission.
For a given cloud temperature, the observed $870\,\mu$m intensities
are then directly proportional to the column densities of the
interstellar matter and thus the line-of-sight extinctions.
These intensities can be converted to 
the beam-averaged hydrogen molecule column density via the formula
\begin{equation} 
N_{\rm H_2} = \frac{F_{\nu}\,R}{B_{\nu}(T_d)\;\Omega\;\kappa_{\nu}\;\mu\,m_{\rm H}}\;\; ,
\end{equation}
where $\Omega$ is the beam solid angle and $\mu$ the mean molecular weight.
Using again the parameters from \citet{Schuller09},
we find the following peak line-of-sight column densities 
for the two locations of maximum cloud emission: 
$N_{\rm H_2} \sim 6 \times 10^{22}\,{\rm cm^{-2}}$ 
(corresponding to a visual extinction of $A_V \sim 65$~mag)
for the cloud near Tr~14,  and
$N_{\rm H_2} \sim 4 \times 10^{22}\,{\rm cm^{-2}}$ ($A_V \sim 50$~mag)
for the cloud at the Treasure Chest Cluster, when assuming cloud
temperatures\footnote{The choice of $T=30$~K was made because
these clouds are irradiated (= heated) by stars either
near or within these clouds.}
of 30~K in both cases.

A very prominent feature in optical images of the CNC
is the {\sf V}-shaped dark cloud with its tip a few arcminutes
south of $\eta$~Car.
The LABOCA map shows that this feature consists of 
a number of different clouds.
Typical intensities in the diffuse parts of these clouds
are in the range $0.2 - 0.5$~Jy/beam, which (assuming again a
typical dust temperature of 30~K)
correspond to column densities of
$N_{\rm H_2} \sim 3 - 7\,\times 10^{21}\,{\rm cm^{-2}}$ 
and extinctions of $A_V \sim 3 - 7$~mag.
These fairly moderate values agree well with the fact that the
{\sf V}-shaped dark cloud is transparent and thus invisible
in near-infrared images of the CNC.


\section{Results for individual parts of the complex}

\subsection{$\eta$ Car}

The famous object $\eta$ Car is a close binary with strong wind-wind
interaction \citep[see, e.g.,][]{Groh10}.
The (sub-)mm emission is known
to be free-free emission from ionized gas in its stellar wind.

In our LABOCA map, we measure a source flux of $\approx 43$~Jy
in a circular aperture of radius $26''$ for $\eta$ Car.
This flux value seen on 26 December 2007
 is nearly identical to the flux derived by \citet{Gomez10}
(42~Jy in a $26''$ radius aperture) from another LABOCA observation
obtained on 23 July 2007.
This observed constancy may be interesting in the context of the 
known strong temporal variability of the (sub-)mm fluxes
by at least a factor of 2 \citep[see, e.g., discussion in][]{Brooks05}
and also with respect to the detection of a recent significant decrease
of the strength of major stellar-wind emission lines in the spectrum 
of $\eta$ Car, which seem to suggest a recent and very rapid decrease 
in the wind density \citep{Mehner10}.

\subsection{Clouds near Tr~16 and the Keyhole Nebula Region}

The loose cluster Tr~16 is located in the
center of the Carina Nebula and includes the optically dominant
star $\eta$~Car as well as the majority of the O-type stars
in the complex.

   \begin{figure}
   \includegraphics[width=9cm]{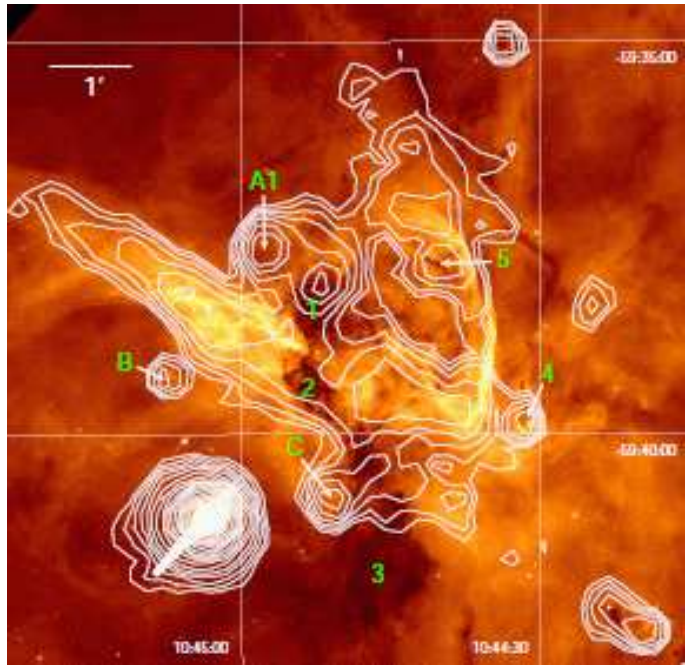}
   \caption{
False-color representation of the
HST H$\alpha$ image of $\eta$~Car and the Keyhole Nebula with
superposed contours of the LABOCA map.
The numbers 1--5, A1, B, and C refer to the CO clumps in the 
Keyhole Nebula defined in \cite{Cox95}.
The contour levels in this image
start at 0.06 Jy/beam and increase on a logarithmic scale
by factors of 1.364 (i.e.~0.135 dex).
              \label{keyhole-laboca-hst}%
    }
    \end{figure}

The Keyhole Nebula Region, immediately to the north of $\eta$~Car,
is the best-observed part of the CNC and the only region
that has already been studied in detail at sub-mm and mm wavelengths.
The comparison of our
LABOCA map to the optical H$\alpha$ mosaic obtained with the
HST is shown in Fig.~\ref{keyhole-laboca-hst}.
Comparison of our LABOCA map to the SIMBA 1.2~mm map of
\citet{Brooks05} generally shows a good agreement in observed
cloud morphologies.
This central region was recently studied with independent LABOCA observations
by \cite{Gomez10}.
With a noise level of $\approx 40$~mJy/beam, their data 
are about half as sensitive as ours.
A comparison of our map to their results shows good agreement
in the general morphology and levels of the sub-mm emission, although our map
reveals (as expected) weaker emission structures.
\cite{Gomez10}
 discuss in detail the relation between the sub-mm emission
and the emission from ionized gas, as seen in H$\alpha$ in the
central few arcminutes around $\eta$~Car.
For most of the cloud structures they found a
close correlation between ionized gas, emission
at both mm and radio wavelengths, and the sub-mm emission.
Together with the clear lack of strong molecular line emission
in this region \citep{Yonekura05} this suggests that
the $870\,\mu$m emission here originates predominantly from 
ionized material,
and not from dust. 

\begin{table}
\caption{LABOCA fluxes of the CO clumps in the Keyhole Nebula listed in \cite{Cox95}}       
\label{table_cocores-kh}      
\centering                          
\begin{tabular}{l r r r r }        
\hline\hline                 
clump &  $M$(CO) & F(870\,$\mu$m) & $M_{870}(20\,{\rm K})$  & $M_{870}(30\,{\rm K})$ \\    
\# & [$M_\odot$] & [Jy] & [$M_\odot$]  & [$M_\odot$] \\    
\hline                        
CB-A1& 14&$0.54\pm0.08$ & 15.8 &  9.1   \\  
CB-B &  3&$0.20\pm0.03$ &  5.8 &  3.4   \\  
CB-C &  6&$0.25\pm0.04$  &  7.3 &  4.2   \\   
CB-1 & 17&       &      &        \\  
CB-2 & 11&       &      &        \\  
CB-3 &  4&$<0.06$&$<1.8$&$<1.0$  \\  
CB-4 &  6& $0.29\pm0.04$ &  8.5 &  4.9   \\   
CB-5 &  1& $0.13\pm0.02$ &  3.8 &  2.2  \\    
\hline                                   
\end{tabular}
\end{table}

However, some weak molecular line emission is nevertheless
present in this region:
\cite{Cox95} discovered CO emission at eight different
positions near the Keyhole Nebula, including  three
peaks aligned with the centers of the optically darkest parts 
of the Keyhole Nebula (their positions 1, 2, and 3). 
In the following discussion of the sub-mm emission seen in our LABOCA map
at the location of these clumps, we use their clump numbers with the prefix ``CB''.
We clearly see compact
$870\,\mu$m emission at the locations of all their clumps with the
exception of CB-1, CB-2, and CB-3.
In order to estimate the fluxes of the sub-mm detected CO clumps, 
which appear as compact sources superposed on top of fairly strong
and inhomogeneous diffuse emission,
we performed aperture photometry at these positions,
using circular apertures with radii of 2.4 pixels and annular sky regions
between $1.5$ times and twice the aperture radius.
Note that the inhomogeneous diffuse background emission
in this area is a serious complication for flux
determinations, and thus the expected uncertainties of our background-subtracted
aperture fluxes listed in Table \ref{table_cocores-kh}
are probably not less than $\sim 30\%$. We also list
clump masses derived via Eq.~(1), assuming two different values for
the temperature, 20~K and 30~K. The mass estimates based on $T=20$~K are
always higher than the masses derived from the CO lines by \cite{Cox95},
but the estimates based on $T=30$~K agree within a factor of $\sim 2$
to the CO masses. This suggests that the material in these clumps
is fairly warm, probably because of the close proximity to the very luminous
stars in this area, which strongly heat them.

For the three CO clumps coinciding with the optically dark Keyhole
nebula features, CB-1, CB-2, and CB-3,
no clear detection of sub-mm emission
can be established from our data.
In deed, we find {\em local minima} of the extended
sub-mm emission in the Keyhole area at these three positions.
The strong $870\,\mu$m emission slightly to the north of CB-1 is
related to the clump CB-A2.
More quantitative statements are hard to make for CB-1 and CB-2,
because these regions are strongly affected by 
blending with the emission
from the long linear structure, Car~II, to the north of $\eta$~Car.
CB-3, on the other hand, is located at a less confused region of our map
and can be characterized in more detail.
 We find no significant sub-mm emission at this position;
with a maximum intensity of 0.04~Jy/beam, all pixel values in this region
remain well below the $3\sigma$ noise limit of 0.06~Jy/beam.
Assuming an upper limit for the sub-mm flux of CB-3 of 0.06~Jy,
we can confirm that given a mass of $4\,M_\odot$ as derived by
\cite{Cox95},  the temperature of this clump must be $\le 12$~K
to explain the non-detection in our sub-mm map.

This cold temperature suggests that this, and the other two
non-detected dark clouds are only seen this close
to the massive stars in projection, but are actually
located at least a few parsec {\em in front} of Tr~16;
otherwise, one would expect considerably
higher cloud temperatures because of the strong radiative heating.

To conclude the discussion of this central region, we note that 
all clouds in this area are only of rather low mass. The original
mass out of which the numerous massive stars formed has been already
nearly completely dispersed, leaving only a few small globules
behind.


\subsection{Clouds near Tr~14}

   \begin{figure}
   \includegraphics[width=9cm]{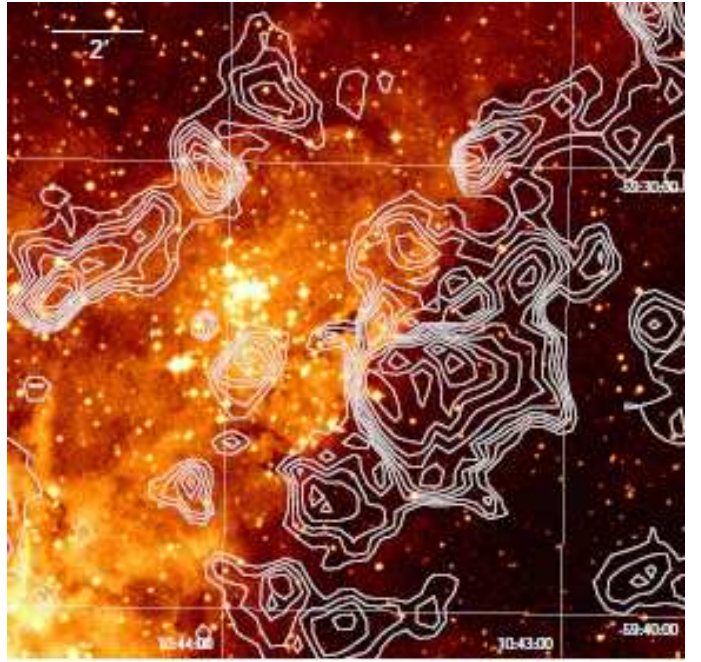}
   \caption{
False-color representation  of the 
red DSS image for the region near Tr~14 with superposed
contours of the LABOCA map. 
              \label{tr14-laboca-dss}%
    }
    \end{figure}

The cluster Tr~14 is the second most massive cluster in the Carina Nebula;
with 10 O-type stars 
it is about a factor of four less massive
than Tr~16, but its spatial configuration is considerably more
compact. 
The brightest sub-mm emitting cloud complex in our map
is found a few arcmin
to the west of Tr~14.
Together with a number of more diffuse clouds to the southeast,
this complex
constitutes the western part of the optically prominent 
{\sf V}-shaped dark nebula in the CNC.

Although the massive cluster Tr~14 contains a substantial number
of hot O-stars that produce large amounts of ionizing flux, the
situation here seems to be quite different from that in the Keyhole
Nebula region. The comparison of the LABOCA maps and the optical image
(Fig.~\ref{tr14-laboca-dss})
clearly shows a strong anti-correlation of H$\alpha$ and sub-mm emission.
This may be at least partly explained by the fact that the cloud
to the west of Tr~14 is much denser and more massive than the
clouds in the Keyhole area, and the ionizing radiation therefore
can affect only the cloud surface, but not penetrate the cloud.
The edge where this cloud faces the stellar clusters is the site
of a prominent Photon Dominated Region (PDR), which has been studied
in some detail by \citet{Brooks03} and \citet{Kramer08}.
From molecular line observations of the dense cloud to the west of Tr~14, 
\citet{Brooks03} estimated 
a total cloud mass of $\sim 20\,000\,M_\odot$ and determined
CO excitation temperatures of $20-30$~K.

Integrating in our LABOCA map over all pixel values above the 
$3\sigma$ noise level
with a radius of $4'$ from the peak of emission, we find a 
$870\,\mu$m flux of 157~Jy.
Most of the flux (98~Jy) comes from the brightest central parts
of the emission at levels above 0.5~Jy/beam.
Assuming a cloud temperature of 20~K, the total cloud
flux of 157~Jy corresponds to 
a cloud mass of $\sim 4600\,M_\odot$,
i.e.~a value substantially lower than the estimate based on 
the molecular line observations.
Considering our above estimate for the column density through
the densest parts of this cloud,
$N_{\rm H_2} \sim 6 \times 10^{22}\,{\rm cm^{-2}}$,
it may well be that the temperature in the inner, dense parts
of the cloud, where most of the mass resides, is lower than 20~K.
Assuming a characteristic dust temperature of 10~K 
yields a mass
of $\sim 14\,400\,M_\odot$. Considering the 
uncertainties in the mass determinations described above,
this value would agree
reasonably well with the mass estimate from CO.
\medskip

Finally, we consider the relation between the stellar cluster
Tr~14 and the surrounding clouds. 
The area within $\la 2'$ of the cluster center is largely
devoid of sub-mm emission, suggesting that the original cloud
has been largely dispersed by the feedback from the massive stars.
In a recent study of the Tr~14 stellar population,
\citet{Sana10} claimed 
an age of no more than $0.3-0.5$~Myr for the cluster.
If this very young age is true, one would expect to see the 
expelled remnants of the original cloud out of which the cluster formed
to be still relatively close to the cluster. Might the elongated cloud
complexes to the south-west and north-east of Tr~14 be this
expelled material? We can investigate this question by comparing
the masses of these clouds to the expected mass of expelled gas in the
formation of a dense cluster.
The total stellar mass of Tr~14 is $M_{\rm stars}\sim 4000\,M_\odot$
\citep{Sana10}.
Assuming that $\la 30\%$ of the original cloud mass have been transformed
into stars
\citep[i.e.~a  star formation efficiency of $\la 30\%$, see][]{Lada03},
and the other $\sim 70\%$ have been expelled, the mass of the expelled
clouds should be $\ga 9300\,M_\odot$.

This estimate is actually consistent with the total mass of the two largest
cloud complexes seen near Tr~14 in Fig.~\ref{tr14-laboca-dss}: 
the total $870\,\mu$m flux of the
elongated cloud northeast of Tr~14 is $\approx 35$~Jy, corresponding to a
cloud mass of $\sim 1000\,M_\odot$ (for an assumed temperature of 20~K).
The total mass of the cloud complex to the southwest of Tr~14 is,
as estimated above, $\sim 14\,000\,M_\odot$.
We thus conclude that the total mass of surrounding clouds within a few
parsecs from Tr~14 is about as large as expected, if we assume these
clouds to be the remnant of the cloud out of which Tr~14 formed and
have been recently expelled by the action of the massive stars in Tr~14.

The spatial distribution of these clouds, however, does not follow
the morphology expected for a (more or less) homogeneous bubble around the cluster,
as seen in many other massive star forming regions \citep[e.g.,][]{Deharveng09}.
Inspection of Fig.~\ref{tr14-laboca-dss} shows no circular
alignment of clouds around Tr~14, but rather a morphology suggesting
a ``broken ring'', which is open in the northwestern and southeastern
direction.
A possible explanation of this morphology is that the clouds form
a thick ring-like structure around Tr~14, which we see nearly edge-on.
\citet{Beaumont10} recently suggested that many apparent bubbles
around young clusters are
actually not three-dimensional spheres, but instead
more or less two-dimensional thick rings. They claim that this
could be the result of star formation and stellar feedback
within a flattened, sheet-like molecular cloud.
The spatial configuration of the clouds around Tr~14 seem to fit
to this model.


\subsection{Clouds near the cluster Tr~15}

The northern cluster Tr~15 is considerably smaller and
less prominent than Tr~14 and Tr~16. Containing 6 O-type stars,
its total ionizing
radiation output is about one order of magnitude
weaker than for Tr~14 and Tr~16.
The cluster is thought to be somewhat older than Tr~14 and
Tr~16, probably around 6--8~Myr
\citep[see, e.g.][]{Tapia03}.

Only moderately bright sub-mm emission is seen in this area
(see Fig.~\ref{laboca-dss}).
None of the clouds near Tr~15
shows morphological indications of feedback from stars in Tr~15.
The cloud to the south-east of the cluster shows a prominent
pillar-like structure, but it points toward the south, 
clearly suggesting that it
is irradiated by massive stars in the south (presumably in Tr~16),
not from Tr~15.

One elongated dusty cloud, pointing toward the southwestern
direction, covers the western half of the cluster area.
It is interesting to note that many of the OB stars in Tr~15
are apparently aligned along the northeastern rim of
this cloud.  This may be an extinction effect, suggesting
perhaps that the cloud is only seen in projection in front of
Tr~15  and is not physically associated.


\subsection{Clouds in the South Pillars}

The so-called South Pillars are a complex of
strongly irradiated clouds south of $\eta$ Car, featuring
very prominently in {\it Spitzer} images.
Our LABOCA map reveals several large and dense, as well 
as numerous small and more diffuse clouds in this area.
A number of these clouds show prominent cometary shapes pointing 
towards $\eta$ Car.
In many cases, the sub-mm morphology closely follows the
shape of the cloud surfaces as seen in the {\it Spitzer} images;
this will be discussed in more detail in Sec.~\ref{spitzer.sec}.
The two northernmost larger clouds in this region constitute
the eastern part of the {\sf V}-shaped dark feature seen in the
optical images.

   \begin{figure}
   \includegraphics[width=9cm]{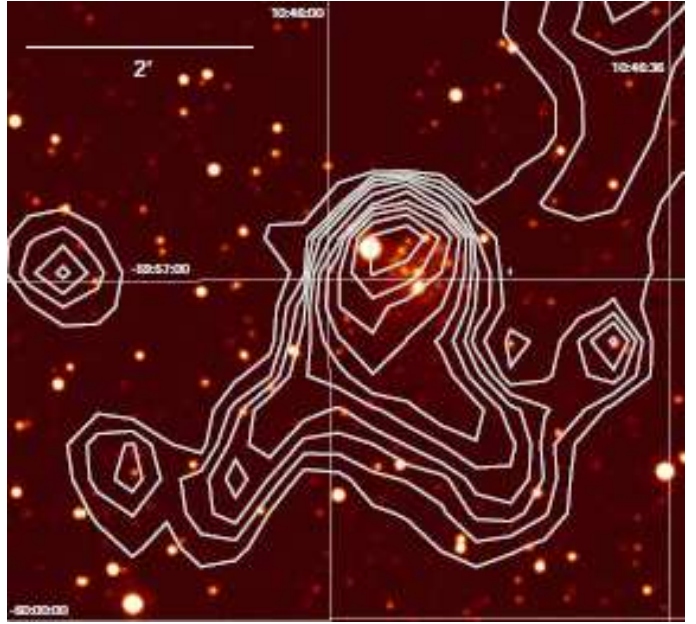}
   \caption{
False-color representation of the
2MASS K-band image of the area around the Treasure Chest Cluster
(located at RA (J2000) = $10^{\rm h}\,45^{\rm m}\,54^{\rm s}$, Dec
= $-59\degr\,56'\,50''$)
with superposed contours of the LABOCA map.
              \label{tcc-opt-2k-lab}%
    }
    \end{figure}

A particularly interesting part of the South Pillars
is the Treasure Chest, a dense
cluster of some 200 stars embedded in a compact nebula at
the head of a large dust pillar \citep{Smith05}.
The brightest object in this cluster is the O9.5~V star 
CPD $-59\degr 2661$, that ionizes a small HII region
inside a $\sim 40''$ diameter cavity near the western edge
of the dust pillar, in which the cluster is located.
The edges of this cavity can be well seen in the narrowband images
presented in \citet[][their Fig.~3]{Smith05}.
The cluster seems to be extremely young ($\la 0.1$~Myr)
and the young stars show strong reddening
with extinction values up to $A_V \sim 50$~mag.

Our LABOCA data reveal the detailed morphology of the
associated dust cloud (see
Fig.~\ref{tcc-opt-2k-lab}) out of which this cluster recently formed.
The large scale structure of the $870\,\mu$m emission
shows a cometary shape,
with peak intensity just to the north and some $15''$ to the
east of the infrared cluster.
At the position of the cavity surrounding the
embedded cluster, the sub-mm emission shows 
a ``hole'' and is $\sim 3 \times$ weaker than on the
eastern side. This suggests that the cavity 
takes up
a large fraction of the full depth of the dust cloud, 
and is not just 
a minor disturbance at the surface of the cloud.

Our column density estimate for the brightest part of the cloud,
$N_{\rm H_2} \sim 4 \times 10^{22}\,{\rm cm^{-2}}$,
corresponds to  visual extinction values of $A_V \sim 50$~mag
and agrees very well with the maximum extinction found by
\citet{Smith05} for the stars in this region.
Proceeding to the south, the 
structure of the cloud shows a bifurcation; the western
arm is denser and thicker than the eastern arm.

To derive an estimate for the total mass of the dust cloud,
we integrated the $870\,\mu$m emission above the $3\sigma$
noise level over a
$\approx 2' \times 3.5'$ box including the brightest part 
of the emission, finding a flux
of 30.2~Jy. Assuming a cloud temperature of 20~K, this 
suggests a cloud mass of $\sim 880\,M_\odot$.


\section{Cloud morphology from the combination of the
{\it Spitzer} and the LABOCA images \label{spitzer.sec}}

   \begin{figure*} \begin{center}
   \includegraphics[width=16.0cm]{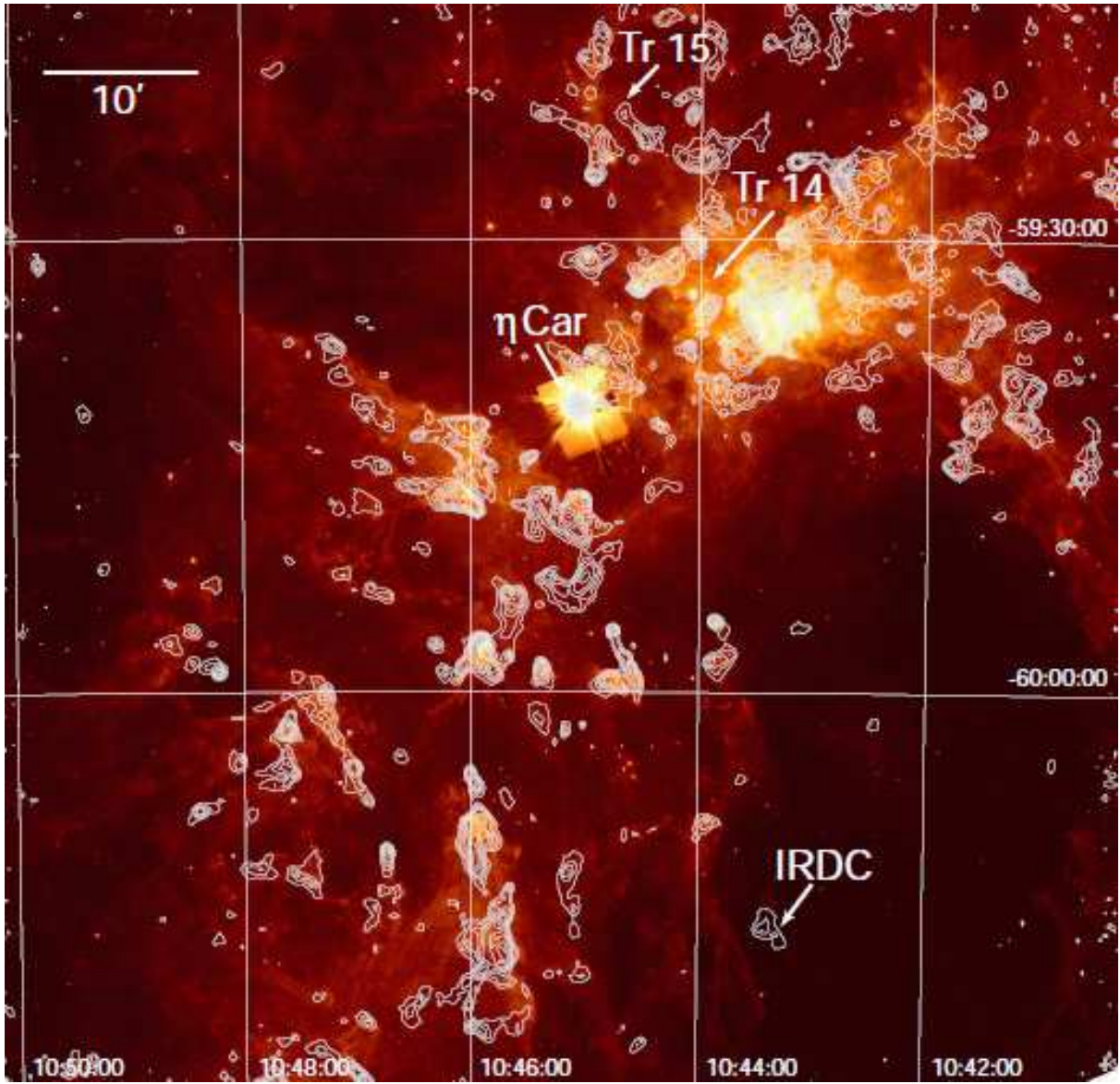} \end{center}
   \caption{
False-color representation of the $8.0\,\mu$m
{\it Spitzer} image 
with superposed contours of the LABOCA map.
In the area within a few arcminutes from $\eta$~Car
the very bright infrared emission 
caused strong artifacts in the {\it Spitzer} image. A grid of J2000
coordinates  is shown.
              \label{sp-s8-lab}%
    }
    \end{figure*}

The combination of our LABOCA data with the existing {\it Spitzer} images
provides a very good way to study the morphology of the clouds.
The $8\,\mu$m emission in the IRAC4 band is dominated by 
polyaromatic hydrocarbon
emission, which is excited by the UV irradiation, and thus
reveals the detailed structure of the cloud surfaces.
The $870\,\mu$m emission, on the other hand, traces the cold
dust inside the dense central cloud regions.
The full extent of the Carina Nebula has been mapped with IRAC,
and we retrieved these data\footnote{These IRAC data were obtained
in July 2008 
in the program ``Galactic Structure and Star Formation in Vela-Carina''
(PI: Steven Majewski, Prog-ID: 40791).}
from the {\it Spitzer} data archive for analysis.
Below, we discuss interesting features resulting
from this investigation.

In Fig.~\ref{sp-s8-lab} we compare the  LABOCA data to the
{\it Spitzer} $8.0\,\mu$m image of the South Pillars region.
In many parts of this region we find a
remarkable resemblance between the morphology in 
the {\it Spitzer} and the LABOCA images.
Essentially all clouds that emit brightly in the {\it Spitzer} image
do also clearly show $870\,\mu$m emission.

\begin{figure*} \sidecaption
  \includegraphics[width=12cm]{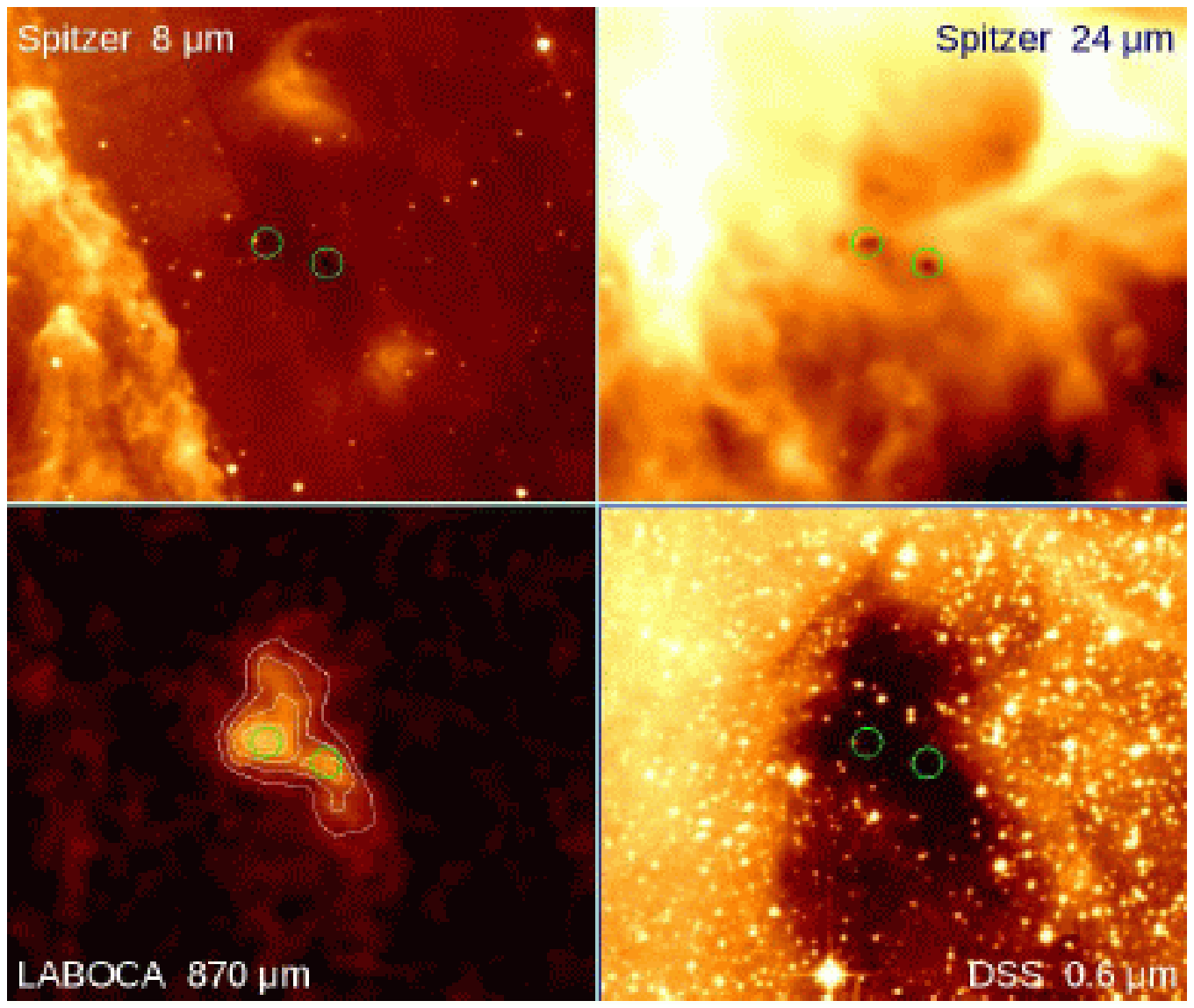}
   \caption{
False-color images of the region around
 the infrared dark cloud, seen in the
{\it Spitzer} $8\,\mu$m IRAC map (upper left), the $24\,\mu$m MIPS map
(upper right), the LABOCA map (lower left), and the optical DSS image
(lower right). s
In each image, the field-of-view is $\approx 8.2'\times 6.6'$,
north is up, east to the left. The contour levels superposed onto the
LABOCA map are 0.06, 0.09, 0.12, and 0.15 Jy/beam. The two peaks
are marked by green circles.
              \label{irdc-mips-lab}%
    }
    \end{figure*}

On the other hand, not all of the dense dusty clouds revealed 
by LABOCA are visible in the {\it Spitzer} images.
An especially notable example is the cloud marked with ``IRDC''
in Fig.~\ref{sp-s8-lab}, near the western edge of the South Pillars.
A more detailed view and comparison to images in other
wavelengths are provided in Fig.~\ref{irdc-mips-lab}.
In the LABOCA map, this cloud shows a double-peaked structure. 
In the {\it Spitzer} $8\,\mu$m IRAC map as well as the $24\,\mu$m MIPS map,
these two $870\,\mu$m peaks correspond to very dark shadows, which are
located at 
R.A.~$= 10^{\rm h}\,43^{\rm m}\,18.2^{\rm s}$,
DEC~$= -60\degr\,15'\,59''$, 
and R.A.~$= 10^{\rm h}\,43^{\rm m}\,24.9^{\rm s}$,
DEC~$= -60\degr\,15'\,40''$.
The western [eastern] peak shows a flux maximum of 0.25~[0.19]~Jy/beam,
which translates into a column density of $N_{\rm H_2} \approx
2~[1.5]~\times 10^{22}\,{\rm cm}^{-2}$, corresponding to a visual extinction
of $A_V \approx 22~[17]$~mag.

The optical DSS image shows that this cloud corresponds 
to a very prominent dark globule in the southern part of the
CNC.
This cloud seems to be a good example of an infrared dark cloud,
i.e.~particularly dense and cold clouds, which are thought to be the
birth places of massive stars \citep[e.g.,][]{Carey98,Rathborne06,Rygl10}.
The integrated 
$870\,\mu$m flux of this cloud is 3.5~Jy, which corresponds to a cloud
mass of $\sim 320\,M_\odot$ for $T=10$~K.

\medskip

\begin{figure}
   \includegraphics[width=9cm]{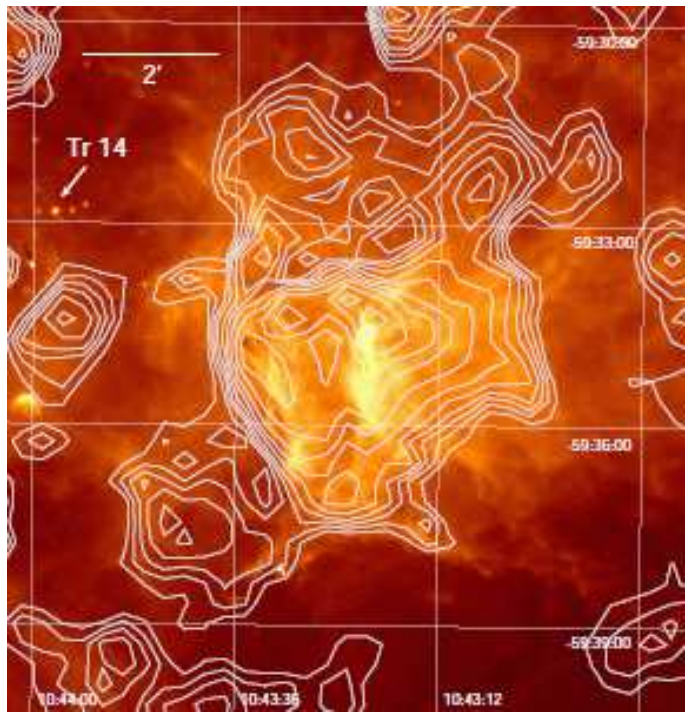}
   \caption{False-color representation of the
$8.0\,\mu$m {\it Spitzer} image  of the cloud west of Tr~14 with
superposed contours of the LABOCA map.
              \label{tr14-sp8-lab}%
    }
    \end{figure}

Finally, in Fig.~\ref{tr14-sp8-lab} we consider the
small scale structure of the massive cloud complex to the west
of Tr~14.
The eastern edge, where the cloud is strongly irradiated by the
massive stars in Tr~14 and where a bright PDR can bee seen
in the optical images, the rim of the cloud 
is very sharp.
The interior of the cloud complex seems to consist of at least
five  bright clumps. To the north and the southeast of this
dense cloud complex,
weaker diffuse emission is visible. This may be
material that was evaporated at the eastern edge of the cloud
in the PDR and is now streaming
away in the directions perpendicular to the impacting
radiation field from Tr~14.

\section{Sub-mm emission from CO molecular cloud clumps}

\cite{Yonekura05} detected 15 individual cloud clumps\footnote{We note
that \cite{Yonekura05} denoted these structures as ``cores'';
however, with a size scale of $\sim 2$ pc, these clouds are not
cores (which have typical sizes of $\la 0.2$~pc) but rather clumps.}
in their
 C$^{18}$O map\footnote{With a half-power beam width of $2.7'$ 
the spatial resolution
of the C$^{18}$O map is about nine times lower than that of our
LABOCA map.}; eight of these (number 8 -- 15) are located
in the field-of-view of our LABOCA map and are marked by the
red circles in Fig.~1.
Sub-mm emission is clearly detected at the locations of all eight
clumps. In five cases (clumps 9, 11, 13, 14, and 15) our LABOCA map
shows strong and compact sub-mm emission, as expected for
high-density clumps. At the location 
of clumps 8, 10, and 12, however, we see only diffuse
sub-mm emission.

\begin{table}
\caption{LABOCA fluxes and mass estimates of the C$^{18}$O clumps detected
by \cite{Yonekura05}}       
\label{table_c18ocores}      
\centering                          
\begin{tabular}{r r r r r r }        
\hline\hline                 
clump & $T_{\rm ex}$ & $M$(C$^{18}$O) & F(870\,$\mu$m) & $M_{870}(20\,{\rm K})$  & $M_{870}(10\,{\rm K})$ \\    
\# & [K] & [$M_\odot$] & [Jy] & [$M_\odot$]  & [$M_\odot$] \\    
\hline                        
8 & 24& 3700&$28.4\pm4.3$ &  829 &  2612   \\  
9 & 21& 1400&$20.3\pm3.0$ &  593 &  1868   \\  
10& 25&  520&$11.9\pm1.8$ &  347 &  1093   \\   
11& 22& 1400&$10.7\pm1.6$ &  312 &   983   \\  
12& 20& 2600&$37.5\pm5.6$ & 1095 &  3450   \\  
13& 20& 4200&$41.5\pm6.2$ & 1212 &  3819   \\  
14&  9&  450&$ 2.9\pm0.4$ &   85 &   267   \\   
15& 19& 2000&$15.5\pm2.3$ &  453 &  1427 \\    
\hline                                   
\end{tabular}
\end{table}

In order to determine the sub-mm fluxes of the clumps, we used
the positions and sizes (the beam-deconvolved radii)
of the clumps reported in Table 3 of
Yonekura and
integrated all pixels values above the $3\sigma$ noise level.
For the clumps with clear peaks in the sub-mm map we
centered the integration regions on these peaks.
The resulting fluxes are given in Table \ref{table_c18ocores}.
We also list the corresponding masses computed from these fluxes,
assuming two different temperature values, $T=20$~K and $T=10$~K.

Comparison to the masses determined
by Yonekura from their  C$^{18}$O map shows that the 
sub-mm mass estimates for $T=20$~K are always substantially
lower than the C$^{18}$O masses. With $T=10$~K we find a much better
agreement; the mass estimates from the sub-mm data and the C$^{18}$O data
generally agree within a factor of $\sim 2$. Although this seems to be
a relatively good match, we note that the assumed very cold $T=10$~K
temperatures appear to be in conflict with the CO excitation temperatures,
which are $\ge 20$~K for all but one of these
clumps. This apparent inconsistency may perhaps indicate that the
derived CO excitation temperatures do not well represent the characteristic
temperature at which most of the cold dust mass resides.


\section{Global properties of the Carina Nebula complex}

Our wide-field LABOCA map yields for the first time
a direct measurement of the total sub-mm emission from the
CNC.
It represents a missing piece of the puzzle for the investigations
of the global properties of the complex, in particular for the
total luminosity and its spectral energy distribution,
and for the total mass  and its distribution in the different
phases, i.e.~dust, molecular, and atomic gas.

\subsection{The total dust and gas mass \label{total_mass_sec}}

We can use our sub-mm map 
to derive an estimate of the total mass of
dust and gas in the Carina Nebula.
We note again that 
our LABOCA map is not sensitive to possibly existing
widespread and spatially homogeneous sub-mm emission on large angular
scales, $> 2.5'$; our measured sub-mm fluxes are therefore only lower limits
to the true flux. 
In principle, if most of the cloud mass is concentrated
into dense compact structures, the 
fraction of unrecovered large-scale flux may be
very small \citep[see, e.g.,][]{Maruta10}.
However, the CNC clearly shows wide-spread
far-infrared emission (as visible, e.g., in the IRAS images) on
scales of tens of arcminutes, and therefore
we expect that the missing flux is not insignificant.
On the other hand, it appears likely that  some part of this
large scale emission
arises from distant clouds in the background;
owing to its position very close to the Galactic plane ($l \approx -0.6\degr$)
and near the tangent point of the Sagittarius-Carina spiral arm, 
the Galactic background
emission at the location of the Carina Nebula must be fairly high.

As determined in Sect.~3, the total flux in our map
originating from dusty clouds (and not from free-free emission)
is  1077~Jy.
To illustrate how sensitively the corresponding mass estimate depends
on the assumed dust temperature, we consider three different 
temperature values, and find that the measured flux
corresponds to (dust + gas) masses of 
$99\,000\,M_\odot$, $31\,400\,M_\odot$, or $18\,100\,M_\odot$
for 10, 20, or 30~K, respectively. 
As our best estimate we assume here a total mass of the cold dusty
clouds traced by LABOCA of $\sim 60\,000\,M_\odot$.

The total mass of {\em molecular gas} in the CNC has been
derived from the CO observations by \citet{Yonekura05}.
Since their full map size is considerably larger than the field-of-view of
our LABOCA map, we added their CO masses for their
sub-regions 1, 2, 3, and 7, which cover
approximately the field of our LABOCA map. This yields 
a total mass of
$141\,000\,M_\odot$ based on their $^{12}$CO map,
$63\,000\,M_\odot$ from their $^{13}$CO map, and
$22\,000\,M_\odot$ from their C$^{18}$O map.
These numbers suggest that a large fraction of the cloud mass resides in 
clouds of moderate density (as traced by $^{12}$CO),
whereas the denser gas (as traced by $^{13}$CO)
and the very dense clouds traced by C$^{18}$O contain
progressively smaller fractions of the total  mass.
Interestingly, our mass estimate for the cloud emission seen by LABOCA
of $\sim 60\,000\,M_\odot$ 
agrees fairly well to the mass in dense, well localized clouds as traced by 
$^{13}$CO. Comparison to the $^{12}$CO mass  suggests that our 
LABOCA map traces about 40\% of the total molecular cloud
emission, a fraction that appears very reasonable considering
the arguments about the unrecovered large-scale sub-mm flux given above.

\subsection{The spectral energy distribution of the complex}

A good way to study the global energetics and properties of the CNC
is to analyze the global spectral energy distribution (SED)
of the complex.
\citet{SB07} constructed the SED in the wavelength range
from  $8\,\mu$m to $100\,\mu$m from
MSX and IRAS data (excluding the flux of the star $\eta$~Car)
and modeled these data (their Fig.~2) with a combination of three different
optically thin graybody components with
discrete temperatures of 220~K, 80~K, and 35~K.
They used this model to estimate the total dust masses associated
to these components and suggested that the CNC contains 
about $10^6\,M_\odot$ of gas; they claimed that 
there is about three times
more gas in atomic form than molecular gas.
However, they also noted 
that their fit is not unique; equally good
fits can be obtained with different sets of temperature components,
and thus these mass estimates are uncertain.

Our LABOCA data provide a very important new SED point
at longer wavelengths, which strongly constrains possible models
of the SED and allows us to obtain new insights into the
mass budget of the complex.
As mentioned above,
the total integrated flux (above the $3\sigma$ noise limit)
in our LABOCA map is 1147~Jy. Subtracting the flux from
$\eta$~Car (43~Jy), to be consistent with the analysis of SB07,
the total flux from the complex is 1100~Jy.
This value is nearly 20 times smaller than the $870\,\mu$m flux predicted
by the the model of SB07 (see their Fig.~2).
Even if we take into account that the field of our LABOCA map
is considerably smaller than the $\simeq 5.6$~square-degree area for which the
SED points in SB07 were determined\footnote{We analyzed the $100\,\mu$m IRAS map
(since the luminosity and mass estimates from SB07 are strongly dominated by
their assumed 35~K model component, the amplitude of which is proportional
to the $100\,\mu$m flux) and found that our LABOCA field encloses 
$\simeq 70\%$ of the total $100\,\mu$m flux in the larger field.},
 a large discrepancy of a factor of $\sim 14$ remains.

Several possible effects could contribute to this discrepancy.
First, the choice of the discrete temperatures for the graybody components
can strongly affect the predicted sub-mm flux; a slightly higher assumed
temperature would reduce the predicted sub-mm flux substantially.
A second aspect is the wavelength-dependence of the dust emissivity:
SB07 assumed an
emissivity~$\propto \lambda^{-1}$, i.e.~a dust emissivity index
 $\beta =1$, which is at the low side
of the range of values found by \cite{Rathborne10} 
for dark clouds ($\beta = 1-2$). Values of $\beta > 1$ would
also reduce the predicted sub-mm flux substantially.
Thirdly, a three temperature model is obviously a simplification,
because the true temperature distribution must be continuous. 
Considering the comparatively small wavelength range covered by the 
 data that were available for their analysis, the use of a 
three temperature model by SB07 was quite appropriate, but the
availability of our new sub-mm data justify a new 
modeling attempt.

\begin{figure}
\includegraphics[width=9cm]{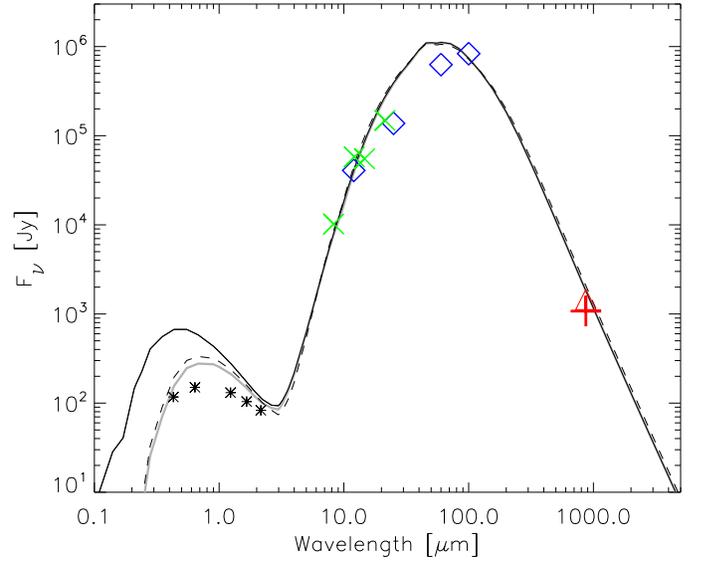}
\caption{Spectral energy distribution of the Carina Nebula complex.
The diamonds and crosses show the mid- and far-infrared fluxes
determined from the IRAS and MSX maps for the field of the LABOCA map.
The cross with the
upward triangle shows the total $870\,\mu$m flux derived from our LABOCA
map (excluding the emission from $\eta$~Car and the Keyhole Nebula).
The asterisks show the integrated optical and near-infrared
fluxes of the known stellar members of Carina
as identified in a deep $Chandra$ X-ray survey \citep[see][]{Preibisch11b}.
The solid line shows the spectral energy distribution resulting
from our spherical cloud model of the CNC with a total gas + dust mass
of $140\,000\,M_\odot$; the gray line shows the effect of adding
1~mag of visual foreground extinction to the spectrum.
The dashed line is the model with a cloud mass of $280\,000\,M_\odot$.
\label{cnc-sed}%
    }
    \end{figure}

\subsection{A simple radiative transfer model of the Carina Nebula complex}

In order to investigate the SED in more detail, we performed
a simple radiative transfer modeling of the
CNC. 
First, we
extracted the fluxes for the SED for the area of our LABOCA map 
from the IRAS\footnote{The total flux in the area
of our LABOCA map was found to be 40\,784~Jy, 137\,539~Jy, 627\,038~Jy, 
and 832\,420~Jy for the wavelengths of $12\,\mu$m,  $25\,\mu$m,
 $60\,\mu$m, and  $100\,\mu$m, respectively.} and MSX images\footnote{In order 
to convert the MSX fluxes, which are given in units of radiance,
to intensity units at the isophotal wavelength, we used the conversion factors
listed in the {\em General Description of MSX Images} from the
NASA/IPAC Infrared Science Archive 
(see {\scriptsize http://irsa.ipac.caltech.edu/applications/MSX/MSX/imageDescriptions.htm}).
For band A, we used an additional correction factor of 2, as recommended
for blackbodies and modified blackbodies at temperatures of about 120--140 K.
The corresponding correction factors for the other bands are very close
to unity. The resulting fluxes are
10\,211~Jy, 57784~Jy, 55200~Jy, and 148479~Jy for
the $8\,\mu$m, $12\,\mu$m, $14\,\mu$m, and  $21\,\mu$m band, respectively.};
the resulting SED is shown in Fig.~\ref{cnc-sed}.

Our radiative transfer modeling of the observed SED  
is not intended to be detailed and highly accurate, but just to
see whether we can reproduce the general shape of the observed
SED with reasonable assumptions about the mass and large-scale
density distribution of the surrounding clouds.
We do not intend to model the small-scale structure of the individual
clouds; instead, we simply assume a central source of radiation
surrounded by a spherical envelope of dust and gas.
Although this is obviously a strong simplification, it provides
the advantage that the temperature distribution
of the gas is computed in a more physically meaningful way
than adding up a few discrete graybody components.

The  radiation transfer calculations were carried out with a
modified version of the code of \cite{Yorke80a,Yorke80b}.
This program yields an exact solution of the frequency-dependent
radiation transfer problem in spherical geometry
simultaneously with a self-consistent determination of the dust 
temperatures.
One has to specify the luminosity and the effective temperature of the
central source, as well as the density distribution of the surrounding cloud 
and can compute the overall spectrum and the flux
distribution at each wavelength.
We use a grid 64 different wavelength points\footnote{The wavelength
grid has a approximately equidistant steps in $\log \lambda$ and
includes most optical to far-infrared standard bands.}
ranging from $0.1\,\mu$m to 5~mm.
The dust model from \cite{Preibisch93} was employed, which had been constructed
to match the
dust properties in molecular clouds. It consists of small
(7--30~nm) amorphous carbon grains and large
(40~nm -- 1~$\mu$m) silicate grains, and assumes that the
silicate grains are coated with a thin mantle (thickness = 14.5\% of the grain
radius) of ``dirty ice'' if their
temperature is below 125~K. The inner and outer
edges of the spatial grid  are at $r = 10^{16}$~cm (0.003~pc)
and 28~pc.

Following the census of massive stars in the CNC by \cite{Smith06},
we assumed a total stellar luminosity of $2.4 \times 10^7\,L_\odot$
and a typical temperature of $T_{\rm eff} = 44\,700$~K.
In the first series of models, we set the 
total (gas + dust) mass of the cloud 
to $140\,000\,M_\odot$, following the CO mass estimate
of \citet{Yonekura05}.
The free parameters in our modeling were the characteristics
of the radial density distribution of the cloud.
The model shown in Fig.~\ref{cnc-sed} assumes that the
density slowly increases with distance from the center
according to $\rho(r) \propto r^{0.75}$; this should approximately
match the conditions in the CNC, where most of the cloud material
has already been dispersed from the central region and is now located
at typical distances between $\sim 5$~pc and $\sim 25$~pc.

The computed dust temperatures in this model
range from 144~K (carbon grains) and 69~K (silicate grains)
at 5~pc, over 103~K and 51~K at 10~pc,
to 34~K and 23~K at 28~pc distance from the central luminosity source.
The spectrum of this simple model
matches the observed mid-infrared to sub-mm fluxes not perfectly, but 
reasonably well.
The predicted $100\,\mu$m flux is 739\,299~Jy, i.e.~$11\%$ lower than
the value extracted
from the IRAS image. The predicted $870\,\mu$m flux is 1816~Jy,
i.e.~$\approx 70\%$
higher than the observed LABOCA flux of 1077~Jy.
This difference agrees well with the expected level of
missing large-scale flux in the LABOCA map estimated above from the
CO masses.  
Assuming the missing flux of 729~Jy to be evenly distributed
over the area of our LABOCA map would require an unrecovered homogenous flux
level of only $\approx 0.014$~Jy/beam.
 
The most important characteristics of this model is that it reproduces
both far-infrared and sub-mm fluxes reasonably well. 
There is therefore
no need to postulate a large mass component of widely distributed
purely atomic gas that would be neither detectable in the CO maps
nor in the sub-mm map.
All available data, i.e.~the far-infrared fluxes, the
sub-mm flux, and the results derived from the CO maps,
can be explained by a total gas and dust mass of about
$140\,000\,M_\odot$.
We note that this number also agrees well with the recent
results from hydrodynamical modeling of the Carina super-bubble
by \citet{Harper-Clark09}, who
derived a total (dust + gas) mass of the nebula of $\sim 10^5\,M_\odot$.
Finally, it is interesting to note that adding 1.0~mag of 
visual foreground extinction
yields quite good agreement between the model spectrum and the
observed integrated optical
and near-infrared fluxes of the stellar member of the CNC.
\smallskip

In a second series of models we investigated the effect of 
increasing the total gas mass above the $140\,000\,M_\odot$ of gas 
as traced by CO emission. In particular, we wanted to find out
whether adding significant amounts
of (atomic) gas would raise the $100\,\mu$m flux to the observed level.
We found this not to be the case. Increasing the total gas mass
actually leads to lower $100\,\mu$m  model fluxes, because the
optical thickness of the gas envelope increases, which leads
to cooler dust temperatures in the outer parts of the nebula.
This effect can be partly compensated for by assuming a steeper outward density
increase. In the second model shown in Fig.~\ref{cnc-sed}
we assumed a total gas mass of $280\,000\,M_\odot$, i.e.~twice the
mass as traced by CO. With a density law $\rho(r) \propto r$
we can find an acceptable SED fit. The predicted $100\,\mu$m flux 
of this model is 722\,721~Jy, 13\% lower than observed.
The $870\,\mu$m flux is 2076~Jy, only 14\% higher than in the
lower mass model; this small increase in sub-mm flux despite the
higher mass is caused by the
cooler dust temperatures in the outer parts of this model
(25~K and 22~K for the carbon and silicate grains at the outer edge of the
model grid).

\subsection{Star formation in the observed clouds}

The dense clouds seen by LABOCA are the site of the current and 
future star-formation activity in the CNC. 
Recent observations have revealed hundreds to thousands of
very young ($\la 1$~Myr old) stellar objects in these clouds
throughout the CNC \citep{Smith10a,Smith10b}.
With our new results on the mass budget of the clouds,
we can address the question of how many additional stars may form 
in the near future in the CNC.

\subsubsection{Cloud mass above the star-formation threshold}

From studies of other molecular cloud complexes it is well
known that typically only a small fraction of the total
cloud mass will form stars, whereas the majority of the
mass will finally be dispersed.
Only the densest (and coldest) parts of the gas in a typical 
molecular cloud will be transformed into stars.
An estimate of the fraction of the cloud mass available for
star formation can be made if the spatial distribution 
of (column) density in the cloud is known.
\citet{Froebrich10} showed that the extinction (or column density)
threshold for
star formation in several nearby molecular clouds is typically 
at $A_V \sim 5 $~mag, corresponding to a column 
density threshold of $N_{\rm H_2} = 4.7 \times 10^{21}\,{\rm cm^{-2}}$.

Assuming a typical cloud temperature of $T=20$~K, we find from our
LABOCA map that the total mass of the clouds above this
column density threshold is about $20\,000\,M_\odot$.
Note that this value agrees quite well with the total mass of 
all C$^{18}$O clumps as determined by \cite{Yonekura05}.
Assuming a total gas mass of $140\,000 - 280\,000\,M_\odot$ (as derived 
above), this implies
that only a small fraction, 7--14\%, of the total gas mass
is in a state 
in which it is available for  star formation.

\subsubsection{The possibility of future massive star-formation in the CNC}

An interesting aspect to investigate is whether 
the current (and future) star-formation process is similar or different
from the previous star-formation activity, which created the massive
star clusters Tr~14, 15, and 16.
A particularly important point in this respect is the
maximum mass of the forming stars.
The clusters Tr~14 and 16 contain numerous
very massive ($M > 50\,M_\odot$) stars.
In the more recently formed embedded population
in the South Pillars, however, the most massive star
identified so far is the O9.5~V star in the Treasure Chest,
which has a comparatively small mass of $\sim 20\,M_\odot$. 
This suggests that
stars in the newest generation have substantially lower maximum
masses than in the older generation.
Will this also be true for the stellar populations yet to form in the
remaining clouds?

Observations of stellar clusters suggest that there
is a relation between the mass $M_{\ast,{\rm max}}$ of the most massive star
in a cluster and the total mass $M_{\rm cluster}$ of all cluster stars
\citep[see][]{Weidner10}.
Although it is still debated whether the observed correlation is actually
caused by physical processes or instead is an effect of
pure random sampling from the IMF in individual star clusters
\citep[see][]{Elmegreen06}, we assume here that 
the simple empirical relation
\begin{equation} M_{\ast,{\rm max}} \approx 1.2\,\times M_{\rm cluster}^{0.45}
\end{equation}
suggested by \citet{Larson03} provides a reasonable approximation.
Assuming a star-formation efficiency (i.e.~the ratio of the
total stellar cluster mass versus the original cloud mass) of ${\rm SFE} 
< 0.3$,
as established for cluster formation by \citet{Lada03}, yields
the desired
relation between the cloud mass and the maximum mass of the stars that
can be expected to form out of this cloud.
It suggests that a cloud with a mass of 
$M_{\rm cloud} \approx 1000\,M_\odot$ 
will yield a maximum stellar mass of $\sim 15\,M_\odot$, whereas
cloud masses of $\ge 12\,000\,M_\odot$ are required
to form a very massive star ($M \ge 50\,M_\odot$).

Nearly all cloud clumps we detected in the CNC have masses $\le 5000\,M_\odot$,
and therefore will according to Eq.~3, yield a maximum stellar mass 
of $\la 30\,M_\odot$.
Most clouds, including the infrared dark cloud discussed above,
have masses $\le 1000\,M_\odot$,
and therefore will yield a maximum stellar mass of just $\la 15\,M_\odot$.
Only the cloud complex to the west of Tr~14 is substantially more
massive. However, as can be seen in Fig.~\ref{tr14-sp8-lab},
this cloud with estimated $\sim 15\,000\,M_\odot$
is already fragmented into several individual clumps; most likely,
each of these individual clumps will form star clusters, and 
none of them is massive enough to form a very massive star ($M \ge 50\,M_\odot$),
if the assumed relation in Eq.~3 holds.

To summarize, the observed cloud masses suggest that the currently
ongoing, presumably triggered \citep[see][]{Smith10b}
star-formation process will probably
{\em not} lead to the formation
of very massive stars, as present in large numbers in the
older, triggering population. This suggests a clear quantitative 
difference in the formation processes of the currently forming and
the earlier generation of stars in the CNC.

\section{Conclusion}

Our wide-field LABOCA map provides the first large-scale
survey of the sub-mm emission in the CNC.
We find that the
cold dust in the complex is distributed in a wide variety
of structures, from the very massive ($\sim 15\,000\,M_\odot$) and
dense cloud complex near Tr~14, over 
several clumps of a few hundred solar masses, to numerous small 
clumps containing only a few solar masses of gas and dust.
Many of the clouds show clear indications that their structure
is shaped by the very strong ionizing radiation field and
possibly stellar winds.

The total mass of the dense clouds to which
LABOCA is sensitive is $\sim 60\,000\,M_\odot$.
This value agrees fairly
well with the mass estimates for the well localized molecular gas traced by
$^{13}$CO. 
The complex may contain a considerable amount 
of very widely distributed atomic gas, which is
neither recovered in our LABOCA map, nor can be seen in the CO data.
Our radiative transfer modeling suggests that the total mass
of such an distributed atomic gas component does probably not
exceed the total molecular gas mass ($140\,000\,M_\odot$) in the region.
Thus, the total gas mass in the field of our LABOCA map seems to  be
$\la 300\,000\,M_\odot$.

Only a small fraction ($\sim 10\%$)  of the gas 
in the CNC is currently in dense and massive enough clouds to be
available for further star formation.
Most observed clouds have masses of less than a few
$1000\,M_\odot$; they will most likely not form
any very massive stars, as present in large numbers in the
older stellar generation in the CNC. This suggests a clear 
quantitative difference between the currently ongoing process
of triggered star-formation and the process that formed
the massive cluster Tr~14, 15, and 16 a few Myr ago.

It is interesting to compare these results about future
star formation in the clouds to the already existing
stellar populations.
The results of a recent wide-field {\it Chandra} X-ray survey
of the CNC \citep{Townsley11} suggest that the total stellar 
population of the
CNC consists of about $43\,000$ stars with a combined mass
of $\sim 27\,000\,M_\odot$ \citep{Preibisch11b}.
This value implies that so far $\sim 10\%$ of the total
cloud mass in the complex have been transformed into stars.
This fraction is similar to the typical values of the
global star-formation efficiency determined for other
OB associations \citep{Briceno07}.
 
With about $\sim 20\,000\,M_\odot$ of dense clouds 
available for star formation at this moment,
the maximum mass of new stars to form within the next $\sim 1$~Myr
may be $\la 7000\,M_\odot$ if we assume a high star formation 
efficiency of 30\%, but probably $\la 2000\,M_\odot$ for lower
star-formation efficiencies.
Compared to the total mass of the already existing stars,
this will be a comparatively small addition. 

The importance of the ongoing and future star-formation 
process in the complex may be increased
by the effect of stellar feedback:
if the irradiation and the winds from the
massive stars efficiently compress the clouds, and 
continuously transform
part of the low-density clouds into denser clouds,
the cloud mass available for star formation could well
increase.
However, it is unlikely that this effect can significantly change
the ratio between the stellar populations,
simply because time is running out:
within less than $\sim 1$~Myr, $\eta$~Car will explode
as a supernovae. This event will be followed by series of 
$\sim 70$ further supernova explosions from the
massive stars in the complex \citep[see][]{SB08}.
Each of these explosions will send strong
shockwaves
through the clouds. While supernova shockwaves are very
destructive for any interstellar material in their immediate
surroundings, they decay into much slower and weaker shocks after traveling
distances of $\ga 1$~pc. Today, most of the molecular clouds in the
CNC are already located at the periphery of the complex,
typically a few~pc away from the massive stars; these clouds
will then be compressed, but probably not destroyed by the
crossing ``evolved'' shockwaves. At
locations where suitable conditions are met 
\citep[see, e.g.,][]{Vanhala98,Oey04},
vigorous star formation activity can then be expected
\cite[see, e.g.,][for the example of the Scorpius-Centaurus Association]{PZ07}.
These supernova shockwaves can
not only trigger star formation but also
inject short-lived radionucleids such as $^{60}$Fe
into the collapsing protostellar clouds \citep{Boss10}.
Because there is strong evidence that such short-lived radionucleids were
incorporated into the solar nebula material during the formation
of our solar system,
the clouds in the CNC may provide a good template in which to 
study the initial conditions for the formation of
our solar system.

\begin{acknowledgements}
We would like to thank the referee for insightful comments 
that helped to improve the paper.
We gratefully acknowledge funding of this work by the German
\emph{Deut\-sche For\-schungs\-ge\-mein\-schaft, DFG\/} project
number PR 569/9-1. Additional support came from funds from the Munich
Cluster of Excellence: ``Origin and Structure of the Universe''.
This publication makes use of data products from the Two Micron All Sky Survey, 
which is a joint project of the University of Massachusetts and the Infrared 
Processing and Analysis Center/California Institute of Technology, funded by 
the National Aeronautics and Space Administration and the 
National Science Foundation.
This work makes use of observations made with the {\it Spitzer} Space Telescope, 
which is operated by the Jet Propulsion Laboratory, California Institute of Technology under a contract with NASA.
The reduction of the HST image used in our analysis was supported by 
STScI Archival Research proposal 11765
(PI Mutchler) and the Hubble Heritage project. 
\end{acknowledgements}


\end{document}